%% file: 0_paper.tex
\documentclass[10pt,conference]{IEEEtran}
\usepackage{amsmath,amsfonts}
\usepackage{algorithmic}
\usepackage{algorithm}
\usepackage{array}
\usepackage[caption=false,font=normalsize,labelfont=sf,textfont=sf]{subfig}
\usepackage{textcomp}
\usepackage{stfloats}
\usepackage{url}
\usepackage{verbatim}
\usepackage{graphicx}
\usepackage{cite}
\usepackage{tabularx}
\usepackage{mdframed}
\usepackage{booktabs}
\usepackage{multirow}
\usepackage{xcolor}
\usepackage{tcolorbox}
\usepackage{xspace}
\usepackage{setspace}
\usepackage{fancyhdr}
\usepackage{lipsum}
\usepackage{import}
\usepackage{caption}  
\usepackage{ragged2e}
 \usepackage{enumitem}
\graphicspath{ {./images/} }
\usepackage{seqsplit}
\usepackage{pifont}

\begin{document}
\input{macros}
\title{
How Effective are Large Language Models in   Generating Software Specifications?
}

\IEEEoverridecommandlockouts

\author{\centering
\IEEEauthorblockN{Danning Xie\IEEEauthorrefmark{1}\IEEEauthorrefmark{3}, Byoungwoo Yoo\IEEEauthorrefmark{2}\IEEEauthorrefmark{3}, Nan Jiang\IEEEauthorrefmark{1}, Mijung Kim\IEEEauthorrefmark{2}\IEEEauthorrefmark{5}, Lin Tan\IEEEauthorrefmark{1}, Xiangyu Zhang\IEEEauthorrefmark{1}, Judy S. Lee\IEEEauthorrefmark{4}}
\IEEEauthorblockA{\IEEEauthorrefmark{1}\textit{Purdue University}, West Lafayette, IN, USA \\ \{xie342, jiang719, lintan\}@purdue.edu, xyzhang@cs.purdue.edu}
\IEEEauthorblockA{\IEEEauthorrefmark{2}\textit{UNIST}, Ulsan, South Korea \\ \{captainnemo9292, mijungk\}@unist.ac.kr}
\IEEEauthorblockA{\IEEEauthorrefmark{4}\textit{ADP}, Roseland, NJ, USA \\ judy.lee@adp.com}
\thanks{\IEEEauthorrefmark{3}The first two authors contributed equally to this paper.}
\thanks{\IEEEauthorrefmark{5}Corresponding author.}
}

\markboth{Journal of \LaTeX\ Class Files,~Vol.~14, No.~8, August~2021}%
{Shell \MakeLowercase{\textit{et al.}}: A Sample Article Using IEEEtran.cls for IEEE Journals}

\IEEEpubid{0000--0000/00\$00.00~\copyright~2021 IEEE}

\maketitle
\begin{abstract}
\input{abstract}

\end{abstract}


\input{1_intro}

\input{background}
\input{2_approach}

\input{experiment}

\input{eval}

\input{future_direction}

\input{threats}

\input{related}

\input{conclusion}
\input{acknowledgment}

\clearpage
\newpage

\bibliographystyle{IEEEtran}
\bibliography{mybib}

\end{document}

%% file: macros.tex


\newcommand{\LineComment}[1]{\Statex /* \textit{#1} */}

\newcommand{\q}[1]{``{#1}''}

\newcommand{\rev}[1]{{#1}}
\newcommand{\bhl}[1]{{\color{red}#1}}
\newcommand{\dep}[1]{{\color{red}#1}}

\newcommand{\mypara}[1]{\vspace{0.03in}\noindent\textbf{#1: }}
\newcommand{\para}[1]{\paragraph{#1}}
\newcommand{\itpara}[1]{\noindent\textit{\textbf{#1 }}}

\newcommand{\dtype}{\textit{dtype}\xspace}
\newcommand{\dtypes}{\textit{dtypes}\xspace}
\newcommand{\shape}{\textit{shape}\xspace}
\newcommand{\datastructure}{\textit{structure}\xspace}
\newcommand{\validvalue}{\textit{valid value}\xspace}

\newcommand{\code}[1]{\texttt{\small#1}\xspace}

\newcommand{\todoc}[2]{{\textcolor{#1}{\textbf{#2}}}}
\newcommand{\todoblack}[1]{{\todoc{black}{\textbf{[[#1]]}}}}
\newcommand{\todored}[1]{{\todoc{red}{\textbf{[[#1]]}}}}
\definecolor{applegreen}{rgb}{0.55, 0.71, 0.0} 
\newcommand{\todogreen}[1]{\todoc{applegreen}{\textbf{[[#1]]}}}
\newcommand{\todoblue}[1]{\todoc{blue}{\textbf{[[#1]]}}}
\newcommand{\todoorange}[1]{\todoc{orange}{\textbf{[[#1]]}}}
\newcommand{\todobrown}[1]{\todoc{brown}{\textbf{[[#1]]}}}
\newcommand{\todogray}[1]{\todoc{gray}{\textbf{[[#1]]}}}
\newcommand{\todopurple}[1]{\todoc{purple}{\textbf{[[#1]]}}}
\newcommand{\todopink}[1]{\todoc{magenta}{\textbf{[[#1]]}}}
\newcommand{\todocyan}[1]{\todoc{cyan}{\textbf{[[#1]]}}}
\newcommand{\todoviolet}[1]{\todoc{violet}{\textbf{[[#1]]}}}
\newcommand{\todoteal}[1]{\todoc{teal}{\textbf{[[#1]]}}}
\newcommand{\todo}[1]{\todored{TODO: #1}}

\definecolor{light-gray}{gray}{0.7}
\newcommand{\hilight}[1]{\colorbox{light-gray}{#1}}

\makeatletter
\newcommand*{\textoverline}[1]{$\overline{\hbox{#1}}\m@th$}
\makeatother

\renewcommand{\todoc}[2]{\relax}

\newcommand{\lin}[1]{\todoblue{Lin: #1}}
\newcommand{\xz}[1]{\todoviolet{Xiangyu: #1}}
\newcommand{\mj}[1]{\todoorange{Mijung: #1}}
\newcommand{\danning}[1]{\todogreen{Danning: #1}}
\newcommand{\byungwoo}[1]{\todobrown{Byungwoo: #1}}
\newcommand{\nan}[1]{\todocyan{Nan: #1}}
\newcommand{\judy}[1]{\todoteal{Judy: #1}}

\newcommand{\docterNumAnno}{2,696}

\newcommand{\jacc}{83.0}
\newcommand{\cacc}{70.0}
\newcommand{\codexJdoctorRandomAccSixty}{87.3}
\newcommand{\gptJdoctorSRAccSixty}{95.6}  

\newcommand{\df}{81.6}
\newcommand{\codexDocTerRandomFSixty}{79.5}
\newcommand{\codexDocTerSRFSixty}{87.2}

\newcommand{\benchmarkCMMSRSixty}{76.4}

\newcounter{finding}
\newcommand{\finding}[1]{\refstepcounter{finding}
 	\vspace{1mm}
	\begin{mdframed}[linecolor=gray,roundcorner=12pt,backgroundcolor=gray!15,linewidth=3pt,innerleftmargin=2pt, leftmargin=0cm,rightmargin=0cm,topline=false,bottomline=false,rightline = false]
		\textbf{Finding \arabic{finding}:} #1
	\end{mdframed}
	\vspace{1mm}
}

\newcommand{\ddata}{DocTer-data\xspace}  
\newcommand{\jdata}{Jdoctor-data\xspace}
\newcommand{\cdata}{CallMeMaybe-data\xspace}
\newcommand{\docter}{DocTer\xspace}
\newcommand{\jdoctor}{Jdoctor\xspace}
\newcommand{\callmemaybe}{CallMeMaybe\xspace}
\newcommand{\basemodel}{CodeLlama-13B\xspace}
\newcommand{\nummodel}{13\xspace}

\captionsetup[figure]{font=bf,skip=6pt}
\captionsetup[table]{font=bf,skip=6pt}
\newcommand{\distance}{8pt}
\setlength{\textfloatsep}{6pt}
\setlength{\floatsep}{\distance}
\setlength{\intextsep}{\distance}
\setlength{\dbltextfloatsep}{\distance} 
\setlength{\dblfloatsep}{\distance} 

%% file: abstract.tex
Software specifications are essential for many Software Engineering (SE) tasks such as bug detection and test generation. Many existing approaches are proposed to extract the specifications defined in natural language form (e.g., comments) into formal machine-readable form (e.g., first-order logic). 
However, existing approaches suffer from limited generalizability and require manual efforts.
The recent emergence of Large Language Models (LLMs), which have been successfully applied to numerous SE tasks, offers a promising avenue for automating this process.
In this paper, we conduct the \textit{first} empirical study to evaluate the capabilities of LLMs for generating software specifications from software comments or documentation. We evaluate LLMs' performance with Few-Shot Learning (FSL) 
and compare the performance of \nummodel state-of-the-art LLMs with traditional approaches on three public datasets.
In addition, we conduct a comparative diagnosis of 
the failure cases from both LLMs and traditional methods, identifying their unique strengths and weaknesses. Our study offers valuable insights for future research to improve specification generation.

%% file: 1_intro.tex
\section{Introduction}

Accurate and comprehensive software specifications are essential for ensuring the correctness, dependability, and quality of software systems~\cite{Blasi2018TranslatingCC, toradocu,Xie2021DocTerDF,Tan2007icommentBO}.
Common software specifications include pre- and post-conditions for a target function that describes the constraints of input parameters and the expected behaviors or output values. They are often required or crucial for 
generating effective test cases and test oracles, symbolic execution, and abnormal behavior identification~\cite{Xie2021DocTerDF, Wong2015DASEDS, korat, cadarKlee}.

Numerous approaches have been proposed to advance automation in extracting specifications from software texts (e.g., documents or comments) into machine-readable forms, including rule-based methods~\cite{Blasi2018TranslatingCC, Tan2012tCommentTJ, Xie2021DocTerDF}, ML-based methods ~\cite{Tan2007icommentBO, pandita2012inferring, callmemaybe}, search-based methods~\cite{Zhai2020C2STN}, etc. 
For example, Jdoctor~\cite{Blasi2018TranslatingCC} leverages pattern, lexical, and semantic matching to translate code comments into machine-readable specifications of pre-/post-conditions, which enables automated test generation that leads to fewer false alarms and the discovery of more defects. 
Several other attempts have been made to further improve these processes in various domains~\cite{Zhai2020C2STN, Lv2020RTFMAA, pandita2012inferring, Zhou2020AutomaticDA}.
However, 
most of existing work is domain-specific, relying on 
heuristics~\cite{Tan2012tCommentTJ, Blasi2018TranslatingCC, callmemaybe} or a large amount of manually 
annotated data~\cite{Xie2021DocTerDF, Zhai2020C2STN}. This reliance makes it challenging to generalize these approaches to other domains.

With the emergence of Large Language Models~(LLMs),
pre-trained on a tremendous amount of documents and source code~\cite{codex, codegen, incoder, codet5, gpt-3, bert}, they have been applied to various Software Engineering (SE) tasks such as code generation~\cite{codegen, liu2023your, yuan2023no, chen2022codet, schafer2023adaptive, ding2024crosscodeeval, zhang2023repocoder} program repair~\cite{clm,zhang2024autocoderover}, and reasoning~\cite{pei2023can}. These models have demonstrated competitive performance compared to traditional approaches~\cite{codegen,codet5,incoder,plbart,clm}. 
Given that software specification extraction predominantly involves the analysis and extraction from software texts, such as comments or documents, and the translation of natural language into (semi-)formal specifications, two research questions naturally arise: 
\textbf{(1) Are LLMs effective in generating software specifications from software texts?}
\textbf{(2) What are the inherent strengths and weaknesses of LLMs for software specification generation compared to traditional approaches?}

\IEEEpubidadjcol
\subsection{Our Study}
To fill in the gap,
we conduct the \textit{first} empirical study to evaluate the capabilities of LLMs in generating software specifications, in comparison with traditional approaches. First, due to the scarcity of labeled data in software specification extraction, we leverage LLMs with {\em Few-Shot Learning}~(FSL)~\cite{Brown2020LanguageMA},
a technique that enables LLMs to generalize from a limited number of examples.
Second, we explore the potential of combining LLMs and FSL by investigating different prompt construction strategies and assessing their effectiveness.
Third, we conduct an in-depth comparative diagnosis of the failure cases from both LLMs and traditional approaches. This allows us to pinpoint their unique strengths and weaknesses, providing valuable insights to guide future research and improvement of LLM applications. Fourth, we conduct extensive experiments, involving \nummodel state-of-the-art LLMs, and evaluate their performance and cost-effectiveness to facilitate model selection in software specification generation. Our study setup and findings are shown below:

\begin{itemize}
    \item \textbf{FSL with random examples outperforms traditional methods:} To assess the performance of LLMs with FSL, we first collect three available datasets from the previous specification extraction research, 
which contain software documents and comments, as well as the corresponding ground-truth specifications.
We start with a basic prompt construction method that \textit{randomly} selects examples for FSL.
We then compare the results with the state-of-the-art specification extraction techniques.
Our 
\underline{Finding 1}
reveals that
\textit{with 10 -- 60 randomly selected examples, LLMs' results 
that are comparable to (2.1\% lower) or better than  (0.8 -- 4.3\% higher) the
state-of-the-art specification extraction techniques. }

\item \textbf{Advanced prompt strategy enlarges the performance gap:} To further explore the potential of LLMs with FSL for specification generation,
we evaluate and compare different prompt construction strategies in terms of their impact on the performance of LLMs. 
Such 
prompt construction 
strategies 
include the above-mentioned random selection
and a \textit{semantics-based} selection strategy.
\underline{Finding 2} shows
that
\textit{with a more sophisticated prompt construction method, the performance gap between LLMs and traditional approaches is enlarged (to 1.9 -- 10.5\%)}.

\item \textbf{LLMs and traditional techniques exhibit unique strengths and weaknesses:}  
While LLMs outperform traditional methods overall, our analysis of failure cases reveals noticeable differences in their failure patterns -- \underline{Finding 3}: \textit{traditional methods often produce empty outputs, whereas LLMs tend to generate incomplete or ill-formed specifications}.
This variation in failures prompts us to analyze the distinct capabilities of LLMs and traditional methods.
We hence conduct an in-depth comparative diagnosis of the failure cases from both ends and investigate their root causes. 
We identified several unique challenges for LLMs, such as \textit{ineffective prompts and missing domain knowledge, which account for 75\% of their unique failures. In contrast, traditional methods fail uniquely 90\% due to insufficient or incorrect rules derived from limited datasets.} (\underline{Findings 4 -- 5})

\item \textbf{Open-sourced \basemodel and StarCoder2-15B are the most competitive models:} Lastly, given the vast spectrum of LLMs in terms of their open-source availability, costs, and model sizes, it becomes imperative to understand their capabilities.
We perform rigorous experiments on \nummodel popular state-of-the-art LLMs, e.g., \basemodel, GPT-4, etc., varying in designs and sizes, and evaluate their performance and cost-effectiveness in generating software specifications. 
Remarkably, 
our \underline{Findings 6 -- 9} show
that \textit{most LLMs achieve better or comparable performance compared to traditional techniques}. \textit{\basemodel and StarCoder2-15B are the overall most competitive model among the \nummodel evaluated models for generating specifications, with high performance,  open-sourced flexibility 
and long prompt support. Their strong performance makes commercial models (e.g., GPT-4) less desirable due to size and cost. }

\item \textbf{Identifying areas for future enhancement:} These findings enable us to identify challenges for further improvement in LLM applications,
i.e.,  
\textit{hybrid approaches}, that integrate LLMs and traditional methods, and \textit{improving prompts effectiveness}. 

\end{itemize}

\subsection{Contributions}
This paper makes the following key contributions:
\begin{enumerate}
    \item We conduct the \textit{first} empirical study
    comparing the effectiveness of LLMs and traditional methods in generating software specifications
    from comments or documents 
    and find that LLMs with FSL 
    achieve results 
    that are comparable to (2.1\% lower) or better than (0.8 -- 4.3\% higher) 
    the traditional methods
    with only 10 -- 60 randomly selected examples (Section~\ref{sec:rq1}).

    \item We evaluate the impact of different prompt construction strategies on the FSL performance and find that the advanced strategy can further enlarge the performance gap between the LLM approach and traditional methods (to 1.9 -- 10.5\%) (Section~\ref{sec:rq2}).

    \item We present a comprehensive failure diagnosis, highlighting \textit{unique} strengths and weaknesses of both traditional methods and LLMs, guiding future research (Section~\ref{sec:rq3}).

    \item We 
    extensively 
    experiment on  \nummodel state-of-the-art  LLMs, assessing their performance and cost-effectiveness (Section~\ref{sec:rq4}).

    \item We discuss the future directions for LLMs in generating software specifications including hybrid approaches and enhanced prompt design~(Section~\ref{sec:futuredirection}).

    \item We release the artifacts in \cite{supplementary}.
    
\end{enumerate}

%% file: background.tex
\section{Background}
\label{sec:background}

\subsubsection{Software Specifications}
\label{sec:background_spec}

Software specifications describe software functionalities, behaviors, and usage, including pre- and post-conditions for functions to ensure correct use. For example, TensorFlow's API \texttt{tf.nn.max\_pool3d} requires the parameter \texttt{input} to be a \q{5-D Tensor}, failing which leads to exceptions. Specifications are critical for tasks like test generation~\cite{Blasi2018TranslatingCC, Xie2021DocTerDF}, program analysis~\cite{Zhai2020C2STN}, bug detection~\cite{Zhou2020AutomaticDA, toradocu, callmemaybe}, and code synthesis~\cite{david2020qsynth}. These are typically from documents or comments in natural language form, requiring extraction into formal machine-readable formats for downstream tasks.

Various approaches~\cite{Tan2007icommentBO, Zhou2020AutomaticDA, Tan2012tCommentTJ, callmemaybe} have been developed to extract such specifications, but they rely on domain-specific heuristics or labeled data, limiting their generalizability. For example, Jdoctor~\cite{Blasi2018TranslatingCC} uses manually crafted patterns with pre-/post-processing, requiring coding and domain expertise.

\subsubsection{Large Language Models (LLMs) and Few-Shot Learning (FSL)}

LLMs, pre-trained on extensive corpora of natural language and code, acquire general knowledge through tasks like masked span and next-token prediction~\cite{bert,Lewis2019BARTDS,gpt-2,gpt-3}. 
To adapt pre-trained LLMs into customized tasks, 
fine-tuning~\cite{bert,T5,clm}, which requires significant labeled data, and prompting~\cite{gpt-3}, which adapts LLMs using task-specific examples without modifying weights, are common approaches.

FSL enhances LLM performance on downstream tasks with limited labeled data~\cite{Brown2020LanguageMA,gpt-3,fsl-nlp1}. In in-context FSL~\cite{gpt-3,in-context}, examples are provided in the input, allowing the model to generalize without altering weights. This makes prompting a practical and efficient method for specification generation.

%% file: 2_approach.tex
\section{Study setup}
\label{sec:study}
\input{images/overview}

Fig.~\ref{fig:overview}  presents an overview of our study. We collect available datasets from previous specification extraction work, containing software documents or comments and the corresponding ground-truth specifications.
We then 
answer four research questions (RQs):

\itpara{RQ1:  How do LLMs with FSL perform compared to traditional rule-based approaches in extracting software specifications?}
We apply 
the benchmark LLM 
to the collected datasets
and compare its performance with the traditional approaches. %
We use a basic prompt construction strategy, {\em random retrieval}, to construct prompts with random examples (Section~\ref{sec:prompt-random}), which coach the LLMs to generate specifications 
 by examples.

\itpara{RQ2: How do prompt construction strategies affect the performance of the LLM approaches?} 
We compare the performance of different prompt construction strategies, i.e., \textit{Random Retrieval} 
and \emph{Semantic Retrieval}. 
Semantic retrieval selects examples based on semantic similarity to the target context (Section~\ref{sec:prompt-sr}).

\itpara{RQ3: What are the unique strengths and weaknesses of LLMs and traditional approaches? }
To provide better insights into the 
capabilities
of different approaches 
and shed light on future research, we conduct a comparative failure diagnosis. 
In particular, we sample a set of cases that the LLM approach succeeds while the traditional approach fails and vice versa. We analyze the  symptoms and root causes  of these failures,  and identify their unique strengths and limitations.

\itpara{RQ4: How do different LLMs compare in terms of their performance and cost for generating software specifications?}
To assess performance and cost-effectiveness,
we conduct extensive experiments with \nummodel state-of-the-art LLMs of different sizes, designs, and so forth.
We employ the best-performing prompt construction strategy (\q{Best Retrieval} in Fig.~\ref{fig:overview}) based on the results of RQ2.

\subsection{Existing 
Specification Extraction and Data}
\label{sec:study-dataset}

\input{tables/dataset}

\input{images/jdcotor_dataset}
\input{images/docter_dataset}
\input{images/callmemaybe_dataset}

We study three 
state-of-the-art 
rule-based approaches for
specification extraction, Jdoctor~\cite{Blasi2018TranslatingCC}, DocTer~\cite{Xie2021DocTerDF}, and \callmemaybe~\cite{callmemaybe},
as well as their respective datasets, containing annotated comments or documents with associated specifications. 
To avoid confusion, we use the terms \jdata, \ddata, and \cdata to refer to the datasets, while \jdoctor, \docter, and \callmemaybe refer to the three approaches.
Table~\ref{tab:dataset} presents the number of data points in each dataset. Rows \q{\#Annotation} lists the number of document-specification pairs annotated

\subsubsection{\jdata and \jdoctor} \jdata contains pre- and post-conditions written as executable Java expressions, translated from Javadoc comments of \texttt{@return}, \texttt{@param}, and \texttt{@throws} tags. Fig.~\ref{fig:jdoctor_dataset} provides an example involving the post-condition for the function \texttt{isNullOrEmpty}. \jdoctor uses a combination of pattern, lexical, and semantic matching to identify key components, such as ⟨subject, predicate⟩ pairs, which are then converted into executable Java expressions through manually defined heuristics.

\subsubsection{\ddata and \docter} 
\label{sec:study-dataset-docter}
\ddata contains DL-specific specifications extracted from the API documents of TensorFlow, PyTorch, and MXNet. Specifications are categorized into four types: \textit{dtype} (data types), \textit{structure} (data structures), \textit{shape} (parameter shape or number of dimensions), and \textit{valid value} (valid ranges or enums). Fig.~\ref{fig:docter_dataset} shows an example data point. DocTer extracts constraints using syntactic rules constructed from annotated API descriptions

\subsubsection{\cdata and \callmemaybe}
\cdata contains temporal constraints represented as event sequences, translated from natural language descriptions in code comments. Fig.~\ref{fig:cmm_dataset} provides an example of a Java function, including its signature, comment, and the translated specifications that capture temporal relations between events  \texttt{this.isEmpty()} and \texttt{this.clear()}. The extraction process of \callmemaybe involves identifying propositions related to temporal dependencies and translating them into temporal constraints using semantic analysis 
and heuristic-based translations.

\subsection{Specification Generation with LLMs}
\label{sec:study-prompt}
\input{images/prompt}

To extract specifications using an LLM, we construct prompts with examples,
i.e., via few-shot learning.

\subsubsection{Few-Shot Learning (FSL)}
\label{sec:study-fsl}

Consider a dataset $D = {(x_i, y_i)}_{i=1}^{|D|}$, where $x$ represents the \textit{context} (e.g., document or comment) and $y$ represents the target software specifications. For each data point $(x_{\mathit{target}}, y_{\mathit{target}})$, we select $K$ examples from the other data points in $D$, excluding the target itself, using leave-one-out cross-validation~\cite{breiman1992submodel,loocv-1,loocv-2}. These $K$ examples, along with $x_{\mathit{target}}$, form the prompt used by the LLM to generate the output $y_{out}$, which is then compared to the ground-truth $y_{\mathit{target}}$.

\subsubsection{FSL Prompt Construction}
\label{sec:prompt_construction}
Fig.~\ref{fig:prompt} 
  shows simplified prompts for each dataset. For \jdata and \cdata, each of the $K$ examples includes a function signature, comment, and the corresponding condition, with the target appended. For \ddata, the prompt includes a function signature, parameter description, and annotated constraints.

We treat the three tag types of \jdata (Table~\ref{tab:dataset}) and the \ddata from three libraries as separate sub-datasets. For example, for a target of \texttt{@param} tag, we only select $K$ examples from the other 242 data points of \texttt{@param} tag, excluding itself. 
Similarly, in \ddata, for a target, we select examples from the same library, excluding itself. 

We study two commonly used strategies to choose the $K$ examples: \textit{random retrieval} and \textit{semantic retrieval}.

\paragraph{Random Retrieval}
\label{sec:prompt-random}
It selects $K$ examples randomly from the dataset, excluding the target, i.e., $D \setminus \{(x_{\mathit{target}}, y_{\mathit{target}})\}$. For instance, with $K=20$, 20 random instances are selected as prompts, excluding the target.

\paragraph{Semantic Retrieval (SR)}
\label{sec:prompt-sr}
It selects examples semantically similar to the target, shown to be more effective than random retrieval~\cite{Rubin2021LearningTR,liu2021makes}. We use RoBERTa-large~\cite{roberta-large}
due to its strong performance on the STS dataset~\cite{reimers-2019-sentence-bert}. The impact of different retrieval models is not studied, as previous research suggests minimal differences in performance~\cite{Shin2021ConstrainedLM}.

\subsubsection{Post-Processing of LLM Output}
\label{sec:strudy-postprocess}
\input{images/equivalence}

For \jdata, LLM completions can be semantically correct but not identical to the annotated specifications. Fig.~\ref{fig:equiv} shows an example where both annotated (highlighted in yellow) and LLM-generated specifications (blue) convey the same condition but differ syntactically. 
Such equivalent but syntactically different specifications frequently occur in LLM results
and makes 
automatic assessment challenging, especially when domain-specific knowledge is needed (e.g., array length is non-negative). Therefore, we manually inspect the generated completions for \jdata and report both the raw accuracy of perfect match and final accuracy after manual corrections. Two authors conducted independent reviews, resolving disagreements with a third.

Note that since \jdoctor uses a pattern-based method, it does not require such post-processing as the output format is constrained by heuristics. 
For \ddata and \cdata, we use perfect match, as post-processing is unnecessary for  \ddata (orderless specifications) and \cdata (deterministic outputs).

\subsection{Studied Large Language Models}
\label{sec:study-models}
\input{tables/models}
Table~\ref{tab:models} summarizes \nummodel LLMs from six series that we study in this paper, 
including the state-of-the-art generic, code-specific, open-sourced, and commercial LLMs. For open-source models, we focus on models that are smaller than 15B due to resource constraints.
We do not focus on models with small sizes (e.g., 1B) as our preliminary experiments show their non-ideal performance. We run open-source models locally using 4 Nvidia RTX A6000 GPUs with 48GB memory.

\subsection{Benchmark LLM}
\label{sec:study-benchmark}
We choose to use \basemodel as the benchmark LLM for the study of RQs 1--3 (Fig.~\ref{fig:overview}). \basemodel is one of the state-of-the-art open-source code LLMs, ensuring the reproducibility of our results.
Additionally, it supports more input tokens 
(Table~\ref{tab:models}), allowing a wide range of experimental settings, such as different numbers of examples in the prompt.

After identifying the best prompt construction strategies (\q{Best Retrieval} in Fig.~\ref{fig:overview}) using \basemodel, we apply it to all other LLMs listed in Table~\ref{tab:models} for the study of RQ4.

%% file: images/overview.tex
\begin{figure}[t]
	\centering
	\includegraphics[width=\linewidth]{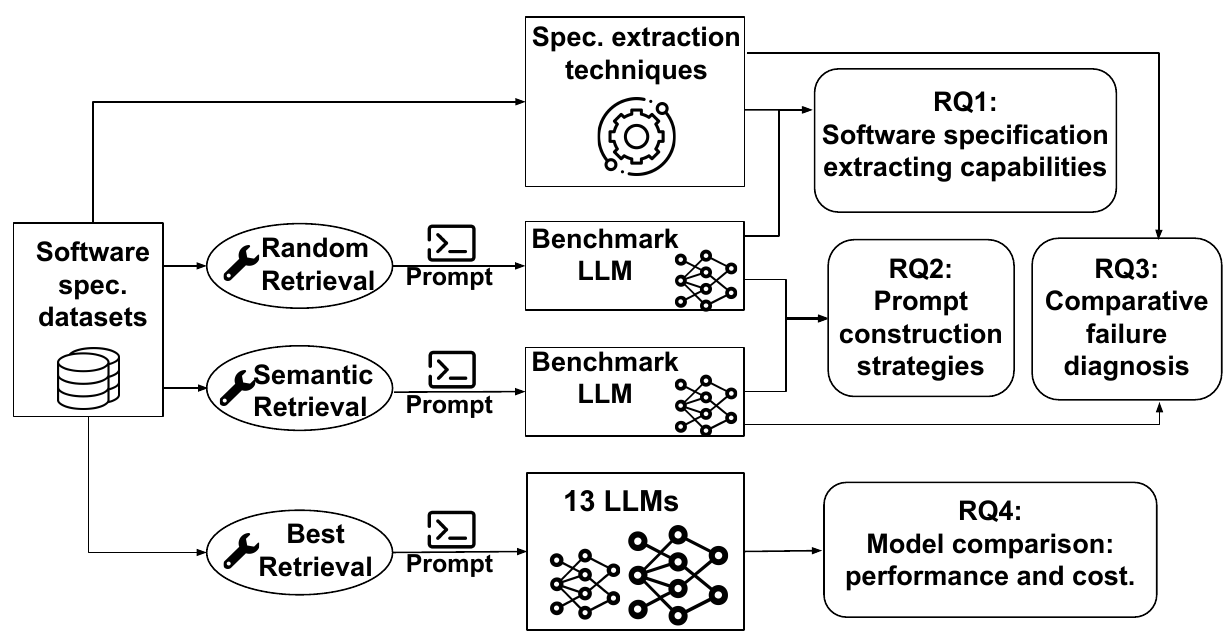}
	\caption{Study overview
	}
	\label{fig:overview}
\end{figure}

%% file: tables/dataset.tex
\begin{table}[t]
\caption{Software specification datasets}
\label{tab:dataset}
\centering
\resizebox{\columnwidth}{!}{%
\begin{tabular}{llllll}
\toprule
\multirow{2}{*}{\jdata~\cite{Blasi2018TranslatingCC}} & Tag Type      & @param     & @return & @throws & Total \\\cline{2-6} 
                         & \#Annotations & 243        & 139      & 472     & 854   \\
                         \midrule
\multirow{2}{*}{\ddata~\cite{Xie2021DocTerDF}}  & Library       & TensorFlow & PyTorch & MXNet   & Total \\\cline{2-6} 
                         & \#Annotations & 1,008        & 484     & 1,384   & 2,876\\
                         \midrule

\cdata~\cite{callmemaybe} &    \#Annotations    &  \multicolumn{3}{c}{89} \\
                        
                         \bottomrule
\end{tabular}%

}

\end{table}

%% file: images/jdcotor_dataset.tex
\begin{figure}[t]
	\centering
	\includegraphics[width=\linewidth]{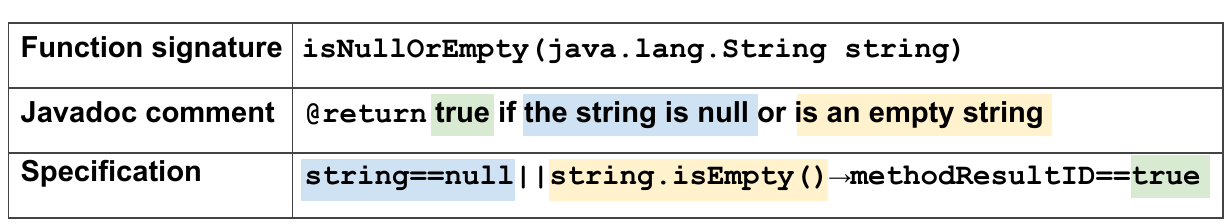}
	\caption{An example data point from \jdata.
 }
	
	\label{fig:jdoctor_dataset}
\end{figure}

%% file: images/docter_dataset.tex
\begin{figure}[t]
	\centering
	\includegraphics[width=\linewidth]{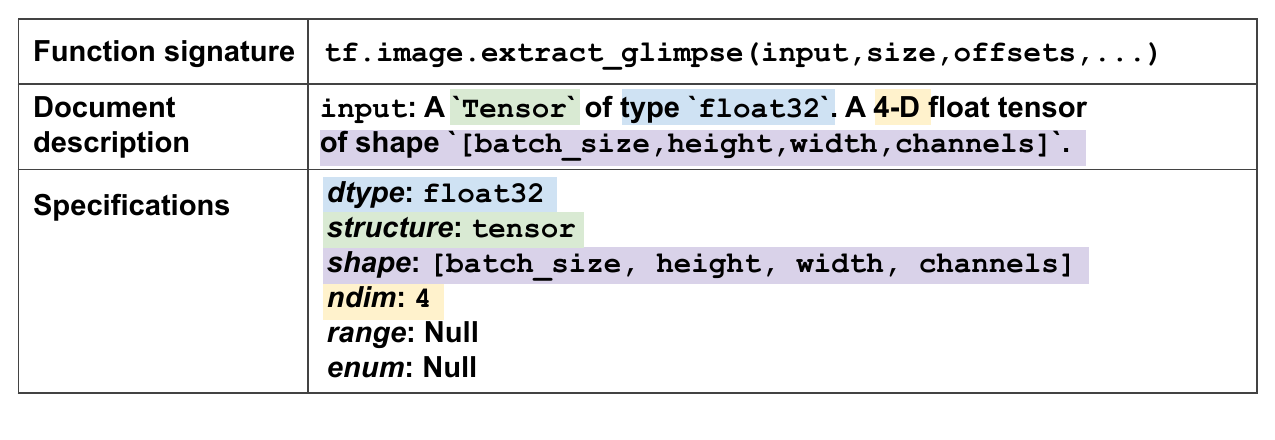}
	\caption{An example data point from \ddata.}
	
	\label{fig:docter_dataset}
\end{figure}

%% file: images/callmemaybe_dataset.tex
\begin{figure}[t]
	\centering
	\includegraphics[width=\linewidth]{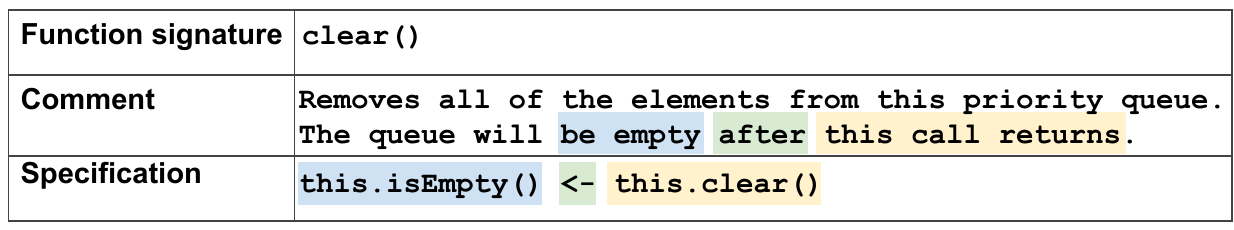}
	\caption{An example data point from \cdata.}
	
	\label{fig:cmm_dataset}
\end{figure}

%% file: images/prompt.tex
\begin{figure}[t]
  \centering
  \subfloat[Jdoctor/CallMeMaybe]{\label{fig:jdoctor_prompt}\includegraphics[clip=true,trim=0mm 0.5mm 0mm 0mm,width=0.49\linewidth]{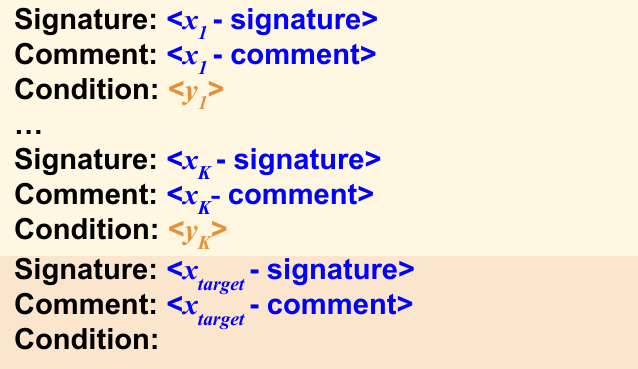}}
  \hfill
  \subfloat[\ddata]{\label{fig:docter_prompt}\includegraphics[clip=true,trim=0mm 0.5mm 0mm 0mm,width=0.49\linewidth]{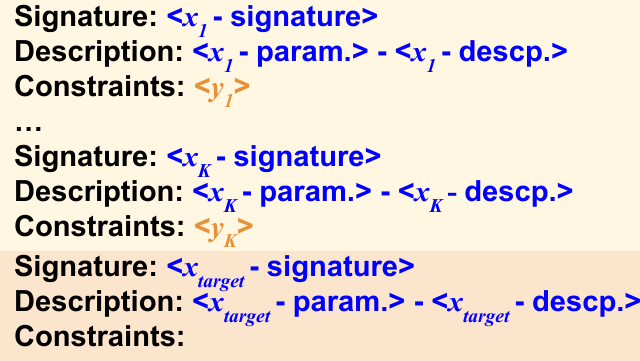}}
  \vspace{-2mm}
  \caption{Prompt structures with target highlighted in orange.
  \danning{ change font; change orange color}\xz{this figure should be made concrete with concrete comments, siganture and so on}}
  \label{fig:prompt}
\end{figure}

%% file: images/equivalence.tex
\begin{figure}[t]
	\centering
	\includegraphics[width=0.7\linewidth]{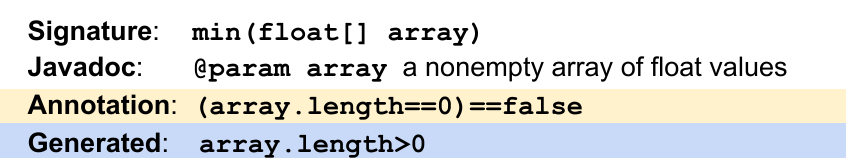}
	\caption{Semantically equivalent specs from \jdata
	}
	\label{fig:equiv}
\end{figure}

%% file: tables/models.tex
\begin{table}[t]

\caption{Studied LLMs:
sizes,
token limits,
prices
per 1,000 tokens,
and open-source statuses.
}
\label{tab:models}
\resizebox{0.95\columnwidth}{!}{%
\begin{tabular}{llclc}
\toprule

Model & \#Param & Token limit & Price (per 1K) & Open-source? \\
\midrule

 GPT4~\cite{gpt-4}& Unknown & 8,192 & \$0.03 &  \ding{55}\\
 \midrule
GPT3.5~\cite{gpt-3.5} & Unknown & 16,384 & \$0.0015 &  \ding{55}\\
\midrule

CodeLlama~\cite{codellama} & 13B, 7B & 16,384 & - & \ding{51} \\

\midrule

Llama3~\cite{llama3} & 8B & 8,192 & - & \ding{51} \\
\midrule
Llama2~\cite{llama2} & 13B, 7B & 4,096 & - & \ding{51} \\

\midrule

deepseek-coder~\cite{guo2024deepseek} & 6.7B & 16,384 & - & \ding{51} \\
\midrule

StarCoder2~\cite{starcoder2}& 15B, 7B & 16,384 & - & \ding{51} \\
\midrule
StarCoder~\cite{li2023starcoder} & 15.5B & 8,192 & - & \ding{51} \\

\midrule

CodeGen2~\cite{nijkamp2023codegen2} & 16B, 7B  & 2,048 & - & \ding{51} \\

 \bottomrule
\end{tabular}%
}
\end{table}

%% file: experiment.tex
\section{Experimental settings}
\label{sec:exp}
\subsection{Model Settings}
\label{sec:exp-model_setting}

As per our experimental design, we truncate examples from the beginning of the prompts to fit the token limit of each model (as shown in Table~\ref{tab:models}). If the median number of tokens in the prompts exceeds the limit, we skip that experiment. For instance, we skip the experiment for  \jdata when $K=60$ for CodeGen2 (with a token limit of 2,048) since the median number of tokens in the prompts is 3,737.
To provide comprehensive results, we run the benchmark model (\basemodel) for RQ1 and RQ2 in all settings.
All models' \texttt{temperatures} are set to 0 for minimal randomness.

\subsection{Accuracy and F1 Metrics}
\label{sec:metrics}
To evaluate the correctness of the generated specifications for \jdata and \cdata, we use accuracy,
defined as the ratio of correctly generated specifications to the total annotated specifications.

For \ddata, we follow the previous work~\cite{Xie2021DocTerDF} and use precision, recall, and F1 to evaluate the generated results for each specification category (e.g., \textit{dtype}). For category $t$, 
let $C_t$ be the number of correctly generated specifications, $N_t$ be the total number of annotated specifications in  the dataset,
and $G_t$ denote the number of generated specifications for category $t$. We define precision as $P_t = \frac{C_t}{G_t}$, recall as $R_t = \frac{C_t}{N_t}$, and F1 score as $F_t = 2 \cdot \frac{P_t \cdot R_t}{P_t + R_t}$.
We report the overall precision, recall, and F1 across all four categories (\textit{dtype}, \textit{structure}, \textit{shape}, and \textit{valid value}) for each library (e.g., TensorFlow).

As discussed in Section~\ref{sec:prompt_construction}, we treat 
\jdata of different tag types and 
\ddata of different libraries as separate datasets. We report the accuracy and F1 metrics for them separately, as well as the overall ones.

%% file: eval.tex
\section{Evaluation  Results}
\label{sec:eval}

\subsection{RQ1: Specification Extracting Capabilities}
\label{sec:rq1}

\input{tables/rq1-jdoctor}

\input{tables/rq1-docter}
\input{tables/rq1-callmemaybe}

We evaluate LLMs on 
\jdata, \ddata, and \cdata
using random retrieval strategy for prompt construction (Section~\ref{sec:prompt-random}) 
with 
\basemodel as the subject model (Section~\ref{sec:study-benchmark}).
Results for both \basemodel and baseline methods (\jdoctor, \docter, and \callmemaybe) are presented in Tables~\ref{tab:rq1-jdoctor}, \ref{tab:rq1-docter}, and \ref{tab:rq1-cmm}. In the tables, we highlight (bold) the results of \basemodel that surpass the baseline methods with the fewest examples used.

As described in Section~\ref{sec:strudy-postprocess}, 
we manually post-process the specification generated for \jdata and present the raw accuracy (automatically calculated with perfect match) in col. \q{Raw} and the final accuracy (after manual correction) in col. \q{Processed} in Table~\ref{tab:rq1-jdoctor}. For \jdoctor (row \textit{\jdoctor}), these two values are the same.
The columns \q{K} 
represent the number of examples in the prompts.

\para{Results Summary} \basemodel demonstrates strong performance across all datasets. 
For \jdata (Table~\ref{tab:rq1-jdoctor}), it achieves 84.0\% accuracy using only 20 randomly chosen examples per comment type, surpassing \jdoctor (\jacc\%). It also outperforms \jdoctor for \texttt{@throws} and matches its performance for \texttt{@param} comments.
For \ddata (Table~\ref{tab:rq1-docter}), it reaches an F1 score of 79.5\% with 60 examples per library, just 2.1\% below \docter (\df\%), despite \docter's need for \docterNumAnno ~annotated examples. 
For \cdata (Table~\ref{tab:rq1-cmm}), \basemodel achieves 70.8\% accuracy with 60 examples, outperforming \callmemaybe by 0.8\%.

\finding{
 \basemodel, with a small number (20 -- 60) of randomly selected examples 
 achieves comparable results (2.1\% lower) with \docter and  
 outperforms the  
 state-of-the-art 
 specification extraction technique \jdoctor and \callmemaybe by 0.8 -- 4.3\%. }

\input{tables/rq2-jdoctor}

\input{tables/rq2-docter}
\input{tables/rq2-callmemaybe}

\input{images/prompt_size}

\subsection{RQ2: Prompt Construction Strategies}
\label{sec:rq2}

Tables~\ref{tab:rq2-jdoctor}, \ref{tab:rq2-docter}, and \ref{tab:rq2-cmm} reveal that \basemodel, when employing the SR strategy, outperforms all baseline methods (\jdoctor, \docter, and \callmemaybe), even with only 10  examples selected in the prompt from each type/category.

Fig.~\ref{fig:promptsize}  
showcases
the effectiveness of random and SR strategies across prompt sizes. SR strategy consistently outperforms both the random strategy and traditional specification techniques across different prompt sizes, highlighting the importance of an appropriate prompt construction strategy for improved FSL performance.

\finding{
The semantic retrieval strategy further improves
\basemodel's performance, leading to a  6.0 -- 10.5\%  improvement over \jdoctor, a 2.7 -- 5.6\% increase over \docter, and a 1.9 -- 6.4\% increase over \callmemaybe. }

\section{RQ3: Comparative Failure Diagnosis}
\label{sec:rq3}

In light of the outstanding performance of LLM compared with traditional techniques, it is crucial to delve deeper into their strengths and limitations.  
We manually examine failing cases of both LLM-based (e.g., \basemodel) and baseline approaches (i.e., \jdoctor,  \docter, and \callmemaybe), and study their failure symptoms and root causes in a comparative manner, aiming to provide insights and directions for future techniques. 
Due to space constraints, we present results for the \jdata dataset, with the other two datasets available in the supplementary material~\cite{supplementary}. The conclusions hold across all datasets, with no significant differences observed.

Fig.~\ref{fig:venn_jdoctor} presents the comparative performance of the \basemodel-based LLM method (L)
and baseline method \jdoctor (J) as a Venn diagram. 
The number in the intersection ($L\cap J$)
denotes cases where both methods are correct, while the number in section $\overline{L\cup J}$
indicates cases they both fail. The presented results are derived from the experiment using \basemodel with SR and $K$ = 60 in RQ2 (Table~\ref{tab:rq2-jdoctor}).

Fig.~\ref{fig:venn_jdoctor} shows that both the LLM and \jdoctor perform well on the majority of cases (80.0\%), indicating that the LLM  quickly learns most specification extraction rules from a small number of examples in the prompts. 
The LLM 
has more (10.7\%) unique correct cases than \jdoctor, indicating the generalizability of LLMs from extensive pretraining. There are a few cases where both methods fail, possibly due to inherent difficulties such as incomplete software text.

To better understand the pros and cons of different methods, we investigate the symptoms and the underlying causes of the failing cases. We \textit{randomly} sample 30 cases from each section of 
Fig.~\ref{fig:venn_jdoctor} 
where at least one of the methods fails,
i.e., $L\cap\overline{J}$, $\overline{L}\cap J$, and $\overline{L\cup J}$.
For the sections that have less than 30 cases (e.g., $\overline{L}\cap J$),
we sample all the failure cases.

The sampling results in 147 failing cases of \basemodel and 135 of the baseline methods across three datasets, with a margin of error of 6\% at a 90\% confidence level.
Two authors categorize these cases independently
with a third author resolving disagreements. 

\subsection{Failure Symptom Analysis}

\input{images/venn_jdoctor}

We conduct further analysis on the distributions of failure symptoms of both LLMs and traditional methods. 
The failure symptoms are classified into four 
categories, \q{ill-formed}, \q{incorrect}, \q{incomplete}, and \q{empty}.
Category \q{\emph{ill-formed}} 
refers to the generated specifications that are invalidly formed.
\q{\emph{Incorrect}} indicates 
specification errors,
while \q{\emph{Incomplete}} denotes the specification is a strict subset of the ground truth. 
\q{\emph{Empty}} denotes a missing specification. The full results and more examples can be found in the artifacts~\cite{supplementary}.

We have the following observations:

\label{sec:rq3-symptom}

\paragraph{LLMs are more likely to generate ill-formed and incomplete specifications}
A small fraction (0--3\%) of \basemodel's failures are ill-formed, unlike traditional rule-based methods that guarantee valid outputs. 
In addition, LLMs are 8\% more likely to produce incomplete specifications compared to rule-based methods, which reliably extract all types by directly matching sequences. In contrast, generative LLMs use sampling to
decode outputs from a distribution, which may occasionally
miss tokens.

\paragraph{Traditional 
techniques are much more likely to generate empty specifications than LLMs} 
For all three datasets, the most common failure of (rule-based) baseline methods is \q{Empty} (67\%), where no specification is generated due to inapplicable rules. 
In contrast, LLMs generate results by predicting missing tokens and will only produce empty results when "empty" is a valid outcome. 
Although 
with a lower empty rate,
LLM tends to generate incomplete and ill-formed specifications as discussed in (a).

\finding{Compared to LLMs, traditional specification extraction approaches are much more likely to generate empty specifications,
while LLMs are more likely to generate ill-formed or incomplete specifications. }

\subsection{Root Cause Analysis}
\label{sec:rq3-rootcause}

\input{images/jdoctor_rq3}

In this section, we categorize the root causes of failures and study their distributions. At the end, we perform a comparative study based on the sections in the Venn diagrams (Fig.~\ref{fig:venn_jdoctor}).

\subsubsection{LLM Failure Root Causes} 
\label{sec:llm_root_cause}
Since LLM results are difficult to interpret, it is in general difficult to determine the root causes of failing cases of LLMs.
We employ the counterfactual method~\cite{rosenbaum2010design, neuberg2003causality}: a structured approach to identifying causal factors by altering one variable at a time to observe the impact on the outcome. 
In particular,
we determine the root cause by finding a fix for it. The nature of the fix indicates the root cause. In some cases, the failure may be fixed in multiple ways. We consider the one requiring the least effort as the root cause and categorize them into five categories.
Fig.~\ref{fig:jdoctor_rootcause}
presents the distributions of the root causes of \basemodel in different sections of the Venn diagrams (Fig.~\ref{fig:venn_jdoctor}). 
We now explain the five categories.

\mypara{Ineffective Prompts}
\input{images/suboptimal_prompt}
It means that the failure is due to the ineffectiveness of the examples in the prompt, even with SR.
Although SR significantly improves FSL's performance (Section~\ref{sec:rq2}), it occasionally falls short in selecting the appropriate examples. 
If we can fix a failing case by manually selecting more relevant example(s) to the prompt, or simply altering the order of the examples in the original prompt, we consider the failure is due to ineffective prompts. 

According to bar \q{\basemodel($\overline{L}$)} from Fig.~\ref{fig:jdoctor_rootcause}, 
26\% \basemodel's failures on \jdata are due to this reason.
We find that the order of examples plays a crucial role, as 21\% of the failure cases in this category are resolved by rearranging the order of the examples.

Fig.~\ref{fig:suboptimal} presents a portion of the 
prompt 
for 
the target parameter \texttt{sp\_input}. 
\basemodel fails to generate the specification \texttt{ndim:>=2}, which is not explicitly stated in the description and requires \basemodel to comprehend the implicit relationship between \texttt{N} and its value range, i.e., \q{N$\ge =$2}. 
Adding an example with such implicit constraints enables \basemodel to generate the correct specification.

\input{images/domain_knolwedge}

\mypara{Missing Domain Knowledge} 
This refers to LLM's failures due to insufficient domain knowledge. For instance, in Fig.~\ref{fig:domainknowledge}, \basemodel generates a specification using a non-existent function,
\texttt{isInDanger}, instead of \texttt{searchForDanger}. 
This issue arises as the LLM lacks relevant context, such as the methods in the class, while \jdoctor 
employs a search-based approach examining all methods in the relevant classes. This result uncovers LLMs' limitation compared to traditional search-based methods: a deficiency in domain knowledge.
To validate our hypothesis,
we manually incorporate relevant domain knowledge into the prompt, alongside the provided examples.
\basemodel then successfully generates the accurate specification, utilizing the correct function (in green).
Fig.~\ref{fig:rootcause} 
shows that 50\% of \basemodel's failures are due to missing domain knowledge.

\mypara{Wrong Focus} 
\input{images/wrong_focus} It denotes instances where LLMs fail to focus on crucial keywords or are misguided or diverted by other content. 
For example, Fig.~\ref{fig:wrong_focus} shows the 
LLM fails to
generate the correct specification from the original document (in yellow), \texttt{numeric}, which specifies the data type of the input tensor.
By employing a slightly revised description that merely quotes the keyword,
the LLM successfully generates the specification.
Fig.~\ref{fig:rootcause} reveals that 13\% of \basemodel's failures are due to the wrong focus.

To identify such failures,
we apply three input mutation strategies: \textit{minor modifications} by simply adding quotation marks to the keywords; \textit{rewriting the sentence} while preserving the same syntactic structure (e.g., changing \q{A or B} to \q{B or A}); 
and \textit{deleting redundant content}
to help the LLM concentrate on the essential parts. All three strategies involve simple and semantics-preserving mutations and do not have impacts on the rule-based methods like \docter
as they rely on syntactic structure. We find that 42\% of such cases can be resolved by merely 
quoting
the keyword(s).

\mypara{Poorly Phrased} It refers to instances where the original documents or comments are ambiguous, poorly written, or hard to understand. Rewriting the sentence to clarify its meaning enables LLMs to generate correct answers. 
According to Fig.~\ref{fig:rootcause}, it contributs to 6\% of \basemodel's failures.

\mypara{Others} We group less common  categories as \q{others}, including \q{contradictory document} and \q{unclear}, accounting for 6\% of \basemodel's failures. 
The former 
refers to buggy or self-contradictory 
comments or documents, causing discrepancies between dataset annotations and LLM-generated specifications.
\q{Unclear} indicates failures with unclear root causes, which we fail to fix despite various attempts.

\finding{
The two dominant root causes combined (ineffective prompts and missing domain knowledge) result in 76\% of \basemodel failures. }

\subsubsection{Baseline Methods Failure Root Causes} 

\label{sec:rq3-baseline-rootcause}

We manually investigated the sampled failing cases for \jdoctor, \docter, and \callmemaybe and identified three root causes: \textit{missing rule}, \textit{incomplete semantic comprehension}, and \textit{incorrect rule}. The distributions on \jdata are in Fig.~\ref{fig:jocter_bl_rootcause}.

\mypara{Missing Rule}
It refers to the absence of relevant rules or patterns, usually resulting in \q{empty}  specifications (Section~\ref{sec:rq3-symptom}). A notable 93\% baseline methods'
failing cases 
fall into this category, exposing a limitation of rule-based methods:
heavily dependent on manually defined or limited rules.

\mypara{Incomplete Semantic Comprehension} 
This occurs when rule-based methods match part of a sentence but fail to grasp its full semantics, leading to incorrect results. For example, \docter extracts a \textit{structure} specification \texttt{vector} from \q{Initializer for the bias vector} but ignores full context or element relationships, affecting the correctness. This accounts for 7\% of failures across the three datasets.

\mypara{Incorrect Rules} 
This category denotes the cases where the applied rules are incorrect. This category of failure is unique to \docter in the three tools we evaluated and is therefore not shown in the figure. 
\docter automatically constructs the rules (map from syntactic patterns to specifications) based on their co-occurrence in the annotated dataset,
which 
can potentially introduce incorrect rules, leading to incorrect extractions. 
10\% of \docter's failing cases are due to \textit{incorrect rules}, while \jdoctor and \callmemaybe does not have any of such failures since their rules are all manually defined.

\subsubsection{Comparative Root Cause Analysis} 
\label{sec:comparative_root_cause}
We compare the root cause distributions of \basemodel (Fig.~\ref{fig:jdoctor_rootcause}), with those of the baseline methods  (Fig.~\ref{fig:jocter_bl_rootcause}).

In cases where the baseline method succeeds (bar \q{\basemodel($\overline{L}\cap J$)}), \basemodel's dominating failure causes are ineffective prompts, missing domain knowledge, and wrong focus. 
Notably, the wrong focus is particularly prevalent here, in contrast to section $\overline{L\cup J}$
where both approaches fail. 
For cases \basemodel succeeds (bar \q{\jdoctor($L\cap \overline{J}$)}) and the baseline method fails,
we observe that the unique baseline failures are primarily due to missing rules and incomplete semantic comprehension. 
That is,
when prompts and software texts are of high quality, \textit{LLMs demonstrate outstanding generalizability},
unbounded by rule sets.
They make predictions based on entire descriptions rather than partial ones.
Notably, 10\% of \docter's unique failures are due to incorrect rules, all of which can be addressed by \basemodel.
We suspect that any automatic rule inference techniques may suffer from such problems if human corrections are not in place.

\finding{
Compared to traditional methods, \basemodel struggles with ineffective prompts and missing domain knowledge, causing 75\% of its \textit{unique} failures. LLMs, however, demonstrate excellent \textit{generalizability}, whereas rule-based approaches often rely on insufficient or incorrect rules extracted from limited datasets.
}

\input{tables/rq3-jdoctor}

\input{tables/rq3-docter}

\input{tables/rq3-callmemaybe}

\subsection{Generalizibility}
\label{sec:rq4:general}
To validate the generalizability of our analysis and conclusions, we extended our evaluation with two additional models, StarCoder-15.5B and GPT-3.5. The distributions from these models align with our findings and conclusions. Detailed results for these models are provided in the artifacts~\cite{supplementary}.

\section{RQ4: Model Comparison} 
\label{sec:rq4}

Tables~\ref{tab:rq3-jdoctor}, \ref{tab:rq3-docter}, and \ref{tab:rq3-cmm}
compare the performance and cost of \nummodel LLMs with the baseline methods (\jdoctor, \docter, and \callmemaybe).
We only list the best results for each model with SR, and the full results can be found in the artifacts~\cite{supplementary}. 
Model response time for generating specifications, subject to various factors such as environment, is omitted. Generally, the response time is reasonably quick for practical usage, ranging between 0.6 to 26.6 seconds.
Due to the token limitation discussed in Section~\ref{sec:exp-model_setting}, some experiments are skipped. 
More experiments on \ddata are skipped since prompts for \ddata are much longer than those for other datasets, making them inapplicable for certain settings.

Overall, generic LLMs with as few as 10 domain examples achieve better or comparable performance as custom-built state-of-the-art specification extraction techniques such as DocTer.
Specifically, 13, 10, and 9 out of the \nummodel models outperform traditional techniques \jdoctor, \docter, and \callmemaybe.

\finding{Most LLMs achieve better or comparable performance as custom-built traditional specification extraction techniques.}

\para{Best Performing Models} Among the \nummodel models, \basemodel and StarCoder2-15B achieve the best performance on all datasets, with an F1 score of 87.9\% (Table~\ref{tab:rq3-docter}) and accuracies of 93.5\% (Table~\ref{tab:rq3-jdoctor}) and 76.4\% (Table~\ref{tab:rq3-cmm}).

\finding{
\basemodel and StarCoder2-15B are the most competitive open-source 
models for extracting specifications, with among the highest performance,
\$0 cost, and long prompt support, facilitating its accessibility, adaptability, and customizability.
}

\para{
Commercial Models (GPT-3.5 and GPT-4)} 

GPT-3.5 and GPT-4  (\$0.0015 and \$0.03 per 1,000 tokens) achieve slightly worse performance than the best-performing models (e.g., \basemodel) given the same number of examples. 
Compared to \basemodel and StarCoder2-15B, which are free, open-source, and with much fewer parameters (Table~\ref{tab:models}), commercial 
models (e.g.,
GPT-4) add no F1 or accuracy gains.
Specifically, GPT-4's  total cost of \$32.8 (for $K=10-60$)
on \cdata causes a 5.6\% accuracy  degradation.

Despite the costs and risks of commercial models, such as charges for usage and concerns regarding the accessibility and continuity of research and applications, they are often more convenient, requiring only an API call with minimal hardware demands. In contrast, open-source models like \basemodel via Hugging Face APIs require more substantial hardware (e.g., GPUs) and technical expertise for configuration and optimization. Both options offer distinct advantages, allowing users to choose based on their needs, resources, and budget.

\finding{
Commercial models (e.g., GPT-4) offer convenience and top-tier performance but with higher costs and risks such as accessibility, whereas open-source models are cost-effective and flexible alternatives but require greater technical expertise and hardware.
}

\para{Other Open-Source Models (CodeLlama-7B, deepseek-coder-6.7B, Llama3, Llama2, StarCoder2-7B, StarCoder, and CodeGen2)}
StarCoder2-7B and CodeLlama-7B offer high performance, slightly below the best-performing model (by 0.7\% -- 1.7\%), presenting it as a viable, smaller alternative to \basemodel and StarCoder2-15B. Conversely, Llama2, despite having the same model sizes as CodeLlama (7B and 13B), performs 3.0\% -- 5.7\% worse, making it less suitable for this task. 
StarCoder-16B and Llama3 demonstrate their strong performance by outperforming the baseline methods by 1.9 -- 10.0\%. CodeGen2, on the other hand, has worse performance on the task. deepseek-coder-6.7B performs 3.7\% worse than \callmemaybe.

\finding{
\basemodel and StarCoder2-15B yield the best performance on software specification generation among tested open-source models. 
Codellama-7B, StarCoder2-7B, StarCoder-16B, and LLama3-8B are reasonable open-source alternatives. 
}

%% file: tables/rq1-jdoctor.tex
\begin{table}[tb!]
\caption{
Comparison of \basemodel with random prompt construction and \jdoctor:  Accuracy (\%).
}
\label{tab:rq1-jdoctor}
\resizebox{\columnwidth}{!}{%
\begin{tabular}{lccccccccc}

\toprule
\multirow{2}{*}{Approach} & \multirow{2}{*}{K} &
\multicolumn{2}{c}{\begin{tabular}[c]{@{}c@{}} 
\texttt{@param}\end{tabular}} & \multicolumn{2}{c}{\begin{tabular}[c]{@{}c@{}}
\texttt{@return}\end{tabular}} & \multicolumn{2}{c}{\begin{tabular}[c]{@{}c@{}}
\texttt{@throws}\end{tabular}} &\multicolumn{2}{c}{\begin{tabular}[c]{@{}c@{}}Overall\end{tabular}}\\ \cline{3-10}

 &  & \multicolumn{1}{c}{Raw} & \multicolumn{1}{c}{Processed} & \multicolumn{1}{c}{Raw} & \multicolumn{1}{c}{Processed} & \multicolumn{1}{c}{Raw} & \multicolumn{1}{c}{Processed} & \multicolumn{1}{c}{Raw} & \multicolumn{1}{c}{Processed} \\
 \midrule
\jdoctor & - & 97.0 & 97.0 & 69.0 & 69.0 & 79.0 & 79.0 & \jacc & \jacc \\
\basemodel & 10 & 81.1 & 89.7 & 35.3 & 48.9 & 76.3 & \textbf{87.3} & 71.0 & 81.7 \\
\basemodel & 20 & 84.4 & 92.6 & 39.6 & 55.4 & 73.9 & 88.0 & 71.3 & \textbf{84.0} \\
\basemodel & 40 & 92.2 & 94.2 & 48.2 & 61.2 & 79.9 & 90.3 & 78.2 & 86.7 \\
\basemodel & 60 & 91.8 & 94.7 & 48.9 & 56.8 & 83.3 & 92.4 & 80.1 & \codexJdoctorRandomAccSixty \\
\bottomrule
\end{tabular}%
}
\end{table}

%% file: tables/rq1-docter.tex
\begin{table}[tb!]
\centering
\caption{Comparison of \basemodel with random prompt construction and \docter: Precision/Recall/F1
(\%).
}
\label{tab:rq1-docter}
\resizebox{\columnwidth}{!}{%
\begin{tabular}{lccccc}
\toprule
Approach & K & \multicolumn{1}{c}{TensorFlow} & \multicolumn{1}{c}{PyTorch} & \multicolumn{1}{c}{MXNet} & \multicolumn{1}{c}{Overall} \\
\midrule
\docter & - & 90.0/74.8/81.7 & 78.4/77.4/77.9 &  87.9/82.4/85.1 & 85.4/78.2/\df \\
\basemodel & 10 &68.3/74.3/71.2 & 72.9/69.1/70.9 & 66.4/71.8/69.0 & 68.2/72.2/70.1 \\
\basemodel & 20 &77.9/72.5/75.1 & 76.6/72.2/74.3 & 71.7/69.7/70.7 & 74.7/71.1/72.8 \\
\basemodel & 40 &77.8/80.4/79.1 & 78.7/72.7/75.5 & 75.3/75.2/75.2 & 76.7/76.6/76.6 \\
\basemodel & 60 & 80.9/79.8/80.4 & 79.6/76.4/\textbf{77.9} & 78.3/80.4/79.4 & 79.4/79.5/\codexDocTerRandomFSixty \\

\bottomrule
\end{tabular}%
}
\end{table}

%% file: tables/rq1-callmemaybe.tex
\begin{table}[tb!]
\centering
\scriptsize
\caption{Comparison of \basemodel with random prompt construction and \callmemaybe: Accuracy (\%).}
\label{tab:rq1-cmm}
\begin{tabular}{llc}
\toprule
Approach    & K  & Accuracy \\
\midrule
\callmemaybe & -  & \cacc     \\
\basemodel   & 10 & 48.3     \\
\basemodel   & 20 & 57.3     \\
\basemodel   & 40 & 62.9     \\
\basemodel   & 60 & \textbf{70.8}    \\
\bottomrule
\end{tabular}
\end{table}

%% file: tables/rq2-jdoctor.tex
\begin{table}[t]
\caption{Comparison of \basemodel using SR prompt construction and Jdoctor:  Accuracy (\%).
}
\label{tab:rq2-jdoctor}
\resizebox{\columnwidth}{!}{%
\begin{tabular}{lccccccccc}
\toprule
\multirow{2}{*}{Approach} & \multirow{2}{*}{K} &
\multicolumn{2}{c}{\begin{tabular}[c]{@{}c@{}}
\texttt{@param}\end{tabular}} & \multicolumn{2}{c}{\begin{tabular}[c]{@{}c@{}}
\texttt{@return}\end{tabular}} & \multicolumn{2}{c}{\begin{tabular}[c]{@{}c@{}}
\texttt{@throws}\end{tabular}} &\multicolumn{2}{c}{\begin{tabular}[c]{@{}c@{}}Overall\end{tabular}}\\ \cline{3-10}

 &  & \multicolumn{1}{c}{Raw} & \multicolumn{1}{c}{Processed} & \multicolumn{1}{c}{Raw} & \multicolumn{1}{c}{Processed} & \multicolumn{1}{c}{Raw} & \multicolumn{1}{c}{Processed} & \multicolumn{1}{c}{Raw} & \multicolumn{1}{c}{Processed} \\
 \midrule
\jdoctor & -                  & 97.0 & 97.0 & 69.0 & 69.0 & 79.0 & 79.0 & \jacc & \jacc \\
\basemodel + \textbf{SR} & 10 & 93.0 & 96.7 & 62.6 & \textbf{71.2} & 84.5 & \textbf{90.3} & 83.4 & \textbf{89.0} \\
\basemodel + \textbf{SR} & 20 & 94.7 & \textbf{97.5} & 63.3 & 73.4 & 86.4 & 91.3 & 85.0 & 90.2 \\
\basemodel + \textbf{SR} & 40 & 95.9 & 97.9 & 62.6 & 72.7 & 89.6 & 94.7 & 87.0 & 82.0 \\
\basemodel + \textbf{SR} & 60 & 95.9 & 98.8 & 65.5 & 75.5 & 90.3 & 96.0 & 87.9& 93.5 \\
\bottomrule
\end{tabular}%
}
\end{table}

%% file: tables/rq2-docter.tex
\begin{table}[]
\centering
\caption{Comparison of \basemodel using SR prompt construction and \docter: Precision/Recall/F1
(\%).
}
\label{tab:rq2-docter}
\resizebox{\columnwidth}{!}{%
\begin{tabular}{lccccc}
\toprule
Approach & K & \multicolumn{1}{c}{TensorFlow} & \multicolumn{1}{c}{PyTorch} & \multicolumn{1}{c}{MXNet} & \multicolumn{1}{c}{Overall} \\
\midrule
\docter & - & 90.0/74.8/81.7 & 78.4/77.4/77.9 &  87.9/82.4/85.1 & 85.4/78.2/\df \\

\basemodel + \textbf{SR} & 10 & 82.8/82.4/\textbf{82.6} & 81.6/80.2/\textbf{80.9} & 86.0/87.4/\textbf{86.7} & 84.1/84.4/\textbf{84.3} \\
\basemodel + \textbf{SR} & 20 & 84.1/84.4/84.2 & 83.6/82.7/83.1 & 86.6/88.4/87.5 & 85.2/86.0/85.6 \\
\basemodel + \textbf{SR} & 40 & 85.9/85.8/85.9 & 83.4/83.4/83.4 & 87.5/88.4/87.9 & 86.2/86.6/86.4 \\
\basemodel + \textbf{SR} & 60 & 86.5/86.3/86.4 & 85.0/84.3/84.6 & 88.3/89.2/88.7 & 87.1/87.4/\codexDocTerSRFSixty \\

\bottomrule
\end{tabular}%
}
\end{table}

%% file: tables/rq2-callmemaybe.tex
\begin{table}[tb!]

\centering
\scriptsize
\caption{Comparison of \basemodel with SR prompt construction and \callmemaybe: Accuracy (\%).}
\label{tab:rq2-cmm}
\begin{tabular}{llc}
\toprule
Approach    & K  & Accuracy \\
\midrule
\callmemaybe & -  & \cacc   \\
\basemodel + \textbf{SR}  & 10 & \textbf{71.9}    \\
\basemodel + \textbf{SR}  & 20 & 70.8     \\
\basemodel + \textbf{SR}  & 40 & 75.3    \\
\basemodel + \textbf{SR}  & 60 & \benchmarkCMMSRSixty    \\
\bottomrule
\end{tabular}
\end{table}

%% file: images/prompt_size.tex
\begin{figure*}[t]
  \centering
  \subfloat[\jdata]{\label{fig:jdoctor_promptsize}\includegraphics[clip=true,trim=0mm 0.5mm 0mm 0mm,width=0.27\linewidth]{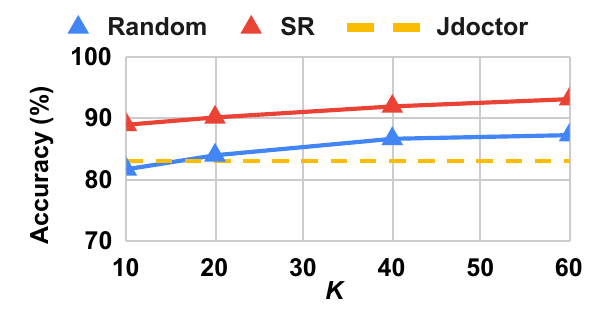}}
  \subfloat[\ddata]{\label{fig:docter_promptsize}\includegraphics[clip=true,trim=0mm 0.5mm 0mm 0mm,width=0.27\linewidth]{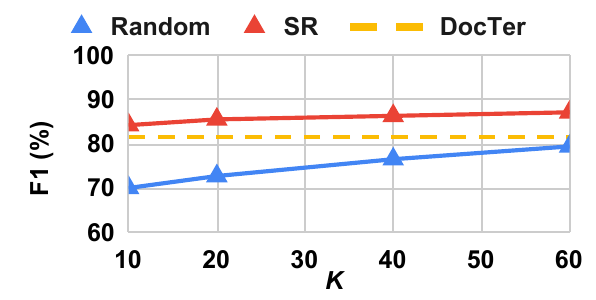}}
    \subfloat[\cdata]{\label{fig:cmm_promptsize}\includegraphics[clip=true,trim=0mm 0.5mm 0mm 0mm,width=0.27\linewidth]{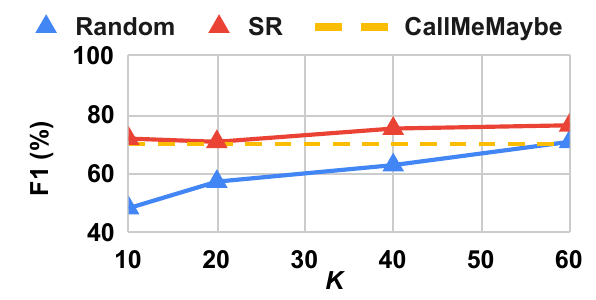}}
  \caption{Comparison of FSL performance using Random and Semantic Retrieval (SR)  for  prompt sizes ($K$) 10 -- 60. 
  }
  \label{fig:promptsize}
\end{figure*}

%% file: images/venn_jdoctor.tex
\begin{figure}[t]
	\centering
	\includegraphics[width=0.5\linewidth]{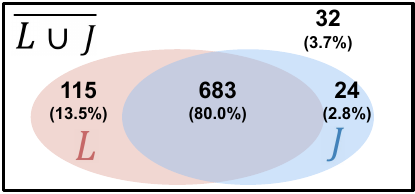}
 \caption{Venn diagrams of specification generation. $L$: \basemodel;  $J$: \jdoctor.
  }
	\label{fig:venn_jdoctor}
\end{figure}

%% file: images/jdoctor_rq3.tex
\begin{figure}[t]
	\centering
	\subfloat[Root causes of \basemodel]{\label{fig:jdoctor_rootcause}\includegraphics[clip=true,trim=0mm 0.5mm 0mm 0mm,width=\linewidth]{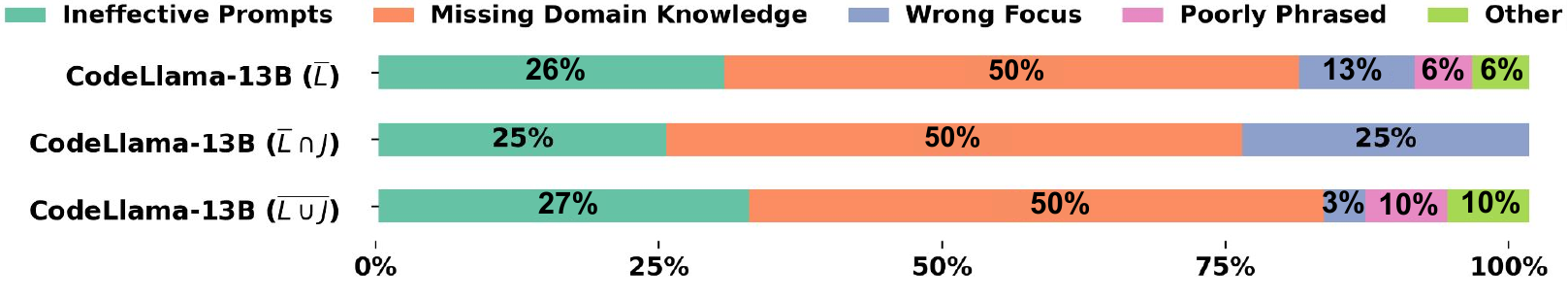}}

	\subfloat[Root causes of the baseline method (Jdoctor)]{\label{fig:jocter_bl_rootcause}\includegraphics[clip=true,trim=0mm 0.5mm 0mm 0mm,width=\linewidth]{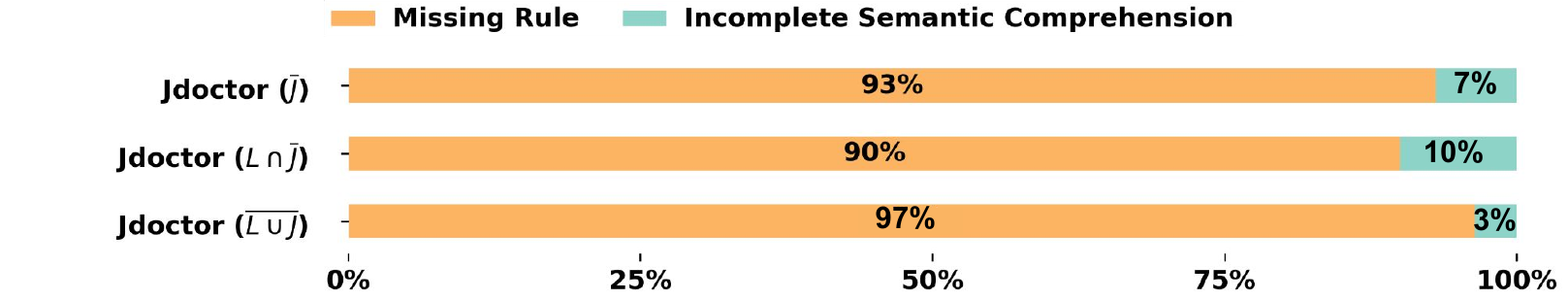}}

	\caption{Distributions of root causes on \jdata failures.}
	\label{fig:rootcause}
   
\end{figure}

%% file: images/suboptimal_prompt.tex
\begin{figure}[t]
	\centering
	\includegraphics[width=\linewidth]{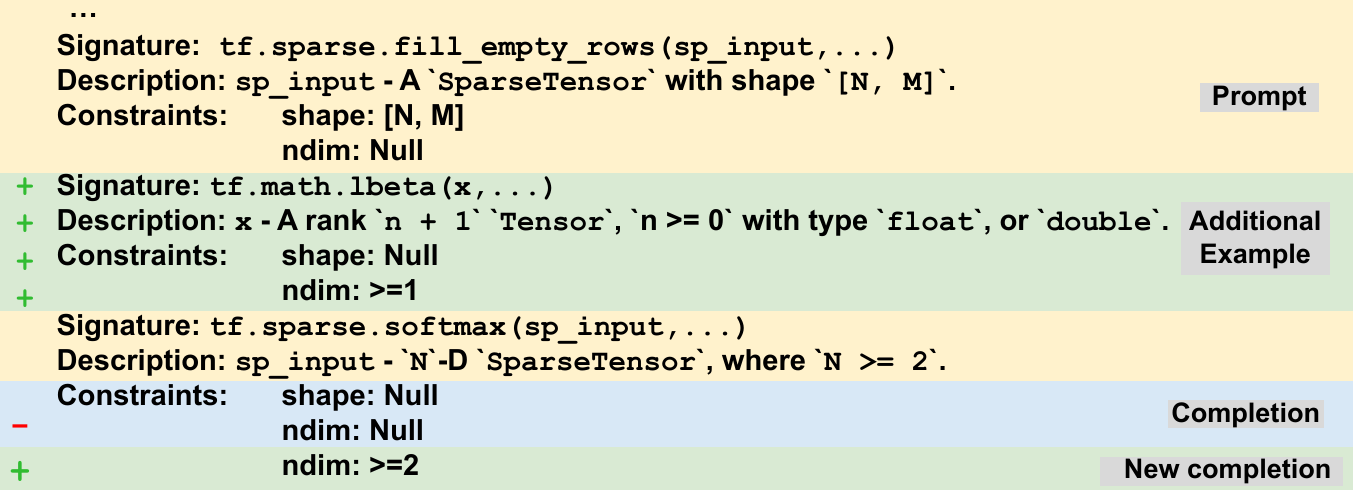}
 \caption{
Example of \q{Ineffective Prompt}.
 Yellow, blue, and green 
 denote
 the original prompt (simplified), generated completion, and an added example that enables LLM to generate the correct specification.
 }

	\label{fig:suboptimal}
\end{figure}

%% file: images/domain_knolwedge.tex
\begin{figure}[t]
	\centering
	\includegraphics[width=\linewidth]{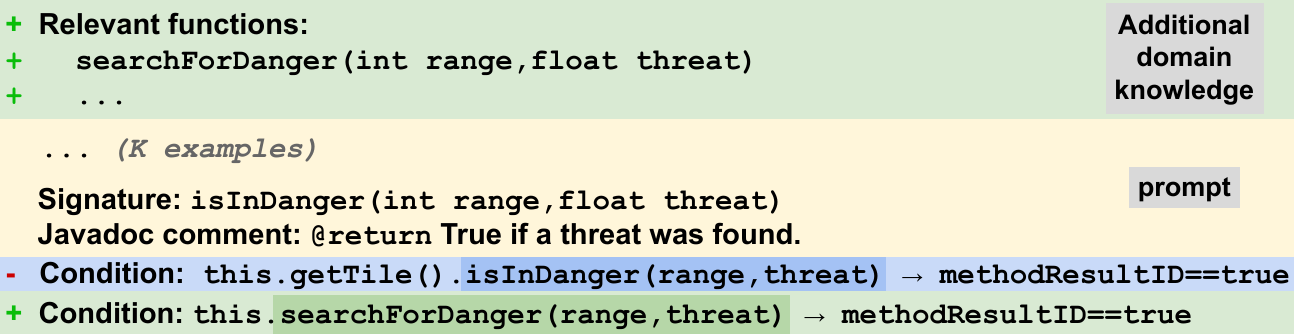}
 \caption{ Example of \q{Missing Domain Knowledge}.
 Yellow, blue, and green 
 denote
 the original prompt
 (simplified),  
 generated completion, added domain knowledge, and the new completion.
 }

	\label{fig:domainknowledge}
\end{figure}

%% file: images/wrong_focus.tex
\begin{figure}[t]
	\centering
	\includegraphics[width=0.5\linewidth]{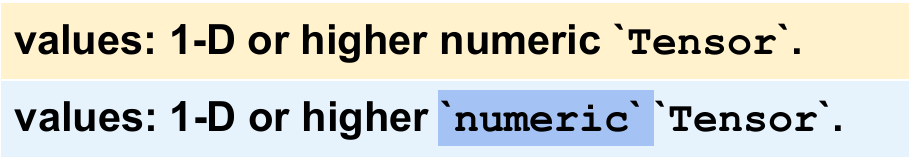}
 \caption{Example of \q{Wrong Focus}.
 A minor adjustment (quoting keyword)   enables 
generating
correct  specification.
 }

	\label{fig:wrong_focus}
\end{figure}

%% file: tables/rq3-jdoctor.tex
\begin{table}[]
\caption{Comparison of different LLMs with SR on \jdata: Accuracy (\%) and Cost (\$).}
\label{tab:rq3-jdoctor}
\resizebox{\columnwidth}{!}{%
\begin{tabular}{lccccccccccc}
\toprule
\multicolumn{2}{c}{\multirow{2}{*}{\begin{tabular}[c]{@{}c@{}}Approach/\\ Model (+SR)\end{tabular}}}  & \multirow{2}{*}{K} &
\multicolumn{2}{c}{\begin{tabular}[c]{@{}c@{}}
\texttt{@param}\end{tabular}} & \multicolumn{2}{c}{\begin{tabular}[c]{@{}c@{}}
\texttt{@return}\end{tabular}} & \multicolumn{2}{c}{\begin{tabular}[c]{@{}c@{}}
\texttt{@throws}\end{tabular}} &\multicolumn{2}{c}{\begin{tabular}[c]{@{}c@{}}Overall\end{tabular}} & 
{\multirow{2}{*}{\begin{tabular}[c]{@{}c@{}}Cost\\ (\$)\end{tabular}}}
\\ \cline{4-11} 

 & & & \multicolumn{1}{c}{Raw} & \multicolumn{1}{c}{Processed} & \multicolumn{1}{c}{Raw} & \multicolumn{1}{c}{Processed} & \multicolumn{1}{c}{Raw} & \multicolumn{1}{c}{Processed} & \multicolumn{1}{c}{Raw} & \multicolumn{1}{c}{Processed} \\

 \midrule
\jdoctor & & - & 97.0 & 97.0 & 69.0 & 69.0 & 79.0 & 79.0 & \jacc & \jacc  & - \\
\midrule

\multirow{2}{*}{GPT} 
& \multirow{1}{*}{4} 
& 60 &  90.1 & 95.5 &  63.3 & 77.7  &90.0   & 94.9  &  85.7 &\textbf{92.3} & 117 \\
& \multirow{1}{*}{3.5} 
& 60 & 86.3 & 89.1 & 59.5 & 72.2 & 80.1 & 84.7 & 79.4 & \textbf{84.4} & 7.8  \\
\midrule

\multirow{2}{*}{Codellama} & \multirow{1}{*}{13B} 
& 60 & 95.9 & 98.8 & 65.5 & 75.5 & 90.3 & 96.0 & 87.9& \fbox{\textbf{93.5}} & -\\
& \multirow{1}{*}{7B} 
& 60 & 95.5 & 98.4 & 60.4 & 72.7 & 90.9 & 95.3 & 87.2 & \textbf{92.5} & -  \\
\midrule

\multirow{1}{*}{deepseek-coder}
 & \multirow{1}{*}{6.7B}
& 60 &  93.8 & 96.3 &  60.4 & 71.2  & 91.5  & 96.0  & 87.1  &\textbf{92.0}  & - \\

\midrule

\multirow{1}{*}{Llama3}
 & \multirow{1}{*}{8B}
& 60 &  95.9 & 96.4 &  65.5 & 77.7  & 90.0  & 94.5  & 87.1  &\textbf{92.9}  & - \\
\midrule 
\multirow{2}{*}{Llama2}
 & \multirow{1}{*}{13B}
& 60 & 95.1 & 98.4 & 56.8 & 68.3 & 88.6 & 93.0 & 85.3 & \textbf{90.5} & - \\
 & \multirow{1}{*}{7B}
& 60 & 95.1 & 97.5 & 56.1 & 67.6 & 86.9 & 91.7 & 84.2 & \textbf{89.4} & - \\


\midrule
\multirow{2}{*}{StarCoder2} & \multirow{1}{*}{15B} 
& 60 & 96.3  &  98.4& 65.0  &77.7& 91.7  &95.6   & 88.7  &\fbox{\textbf{93.5}} & - \\
& \multirow{1}{*}{7B} 
& 60 &  95.9 & 97.9 &  64.0 & 76.3  &90.5   & 94.5  & 87.7  &\textbf{92.5}  & - \\

\midrule
\multirow{1}{*}{StarCoder} & \multirow{1}{*}{16B} 
& 60 & 95.9 & 98.4 & 64.7 & 77.7 & 90.3 & 94.7 & 87.7 & \textbf{93.0} & - \\

\midrule

\multirow{2}{*}{CodeGen2}
& \multirow{1}{*}{16B} 
& 20 & 93.4 & 98.9 & 60.8 & 70.9 & 84.2 & 88.1 & 84.0 & \textbf{89.0} &- \\
 & \multirow{1}{*}{7B} 
& 20 & 91.8 & 96.7 & 64.6 & 72.2 & 84.5 & 87.6 & 84.2 & \textbf{88.3} & -  \\

\bottomrule
\end{tabular}%
}
\end{table}

%% file: tables/rq3-docter.tex
\begin{table}[]
\caption{Comparison of different LLMs with SR on \ddata: Precision/Recall/F1 (\%), and Cost (\$).  }
\label{tab:rq3-docter}
\resizebox{\columnwidth}{!}{%
\begin{tabular}{lccccccccc}
\toprule

\multicolumn{2}{c}{Approach / Model (+SR)} & K & \multicolumn{1}{c}{TensorFlow} & \multicolumn{1}{c}{PyTorch} & \multicolumn{1}{c}{MXNet} & \multicolumn{1}{c}{Overall} & \multicolumn{1}{c}{Cost (\$)} \\
\midrule

 \docter &  & - & 90.0/74.8/81.7 & 78.4/77.4/77.9 &  87.9/82.4/85.1 & 85.4/78.2/\df & - \\

\midrule

\multirow{2}{*}{GPT} 
 & \multirow{1}{*}{4} 
 & 60 & 84.7/87.5/86.1 & 82.8/87.3/85.0 & 86.8/89.9/88.3 & 85.4/88.6/\textbf{87.0} & 78 \\
 & \multirow{1}{*}{3.5} 
 & 20 & 81.9/83.7/82.8 & 77.3/81.2/79.2 & 84.3/87.3/85.8 & 82.3/85.0/\textbf{83.6} & 5.2 \\

\midrule

\multirow{2}{*}{CodeLlama} & \multirow{1}{*}{13B} 
& 60 & 86.5/86.3/86.4 & 85.0/84.3/84.6 & 88.3/89.2/88.7 & 87.1/87.4/\textbf{\codexDocTerSRFSixty}  & - \\
& \multirow{1}{*}{7B} 
 & 60 & 86.4/83.6/85.0 & 84.4/80.4/82.4 & 88.2/88.8/88.5 & 86.9/85.6/\textbf{86.2} & - \\
\midrule

 deepseek-coder & 6.7B & 60 &  85.9/85.4/85.7 & 85.6/84.6/85.1 & 87.5/89.1/88.3 & 86.6/87.0/\textbf{86.9} & - \\
\midrule

 Llama3 & 8B & 60 & 86.1/82.8/84.4 & 83.3/83.0/83.1 & 87.7/85.6/86.7 & 86.4/84.2/\textbf{85.3}  & - \\
\midrule
\multirow{2}{*}{Llama2}
 & \multirow{1}{*}{13B}
& 20 & 83.9/80.0/81.9 & 80.4/74.2/77.2 & 84.4/81.3/82.8 & 83.6/79.6/81.5 & - \\

& \multirow{1}{*}{7B}
& 20 &  80.5/79.2/79.8 & 77.7/75.5/76.6 & 84.0/80.6/82.3 & 81.7/79.3/80.5 & - \\
\midrule

\multirow{2}{*}{StarCoder2} & \multirow{1}{*}{15B} 
& 60&  87.9/87.2/87.6 & 85.8/85.4/85.6 & 88.6/89.4/89.0 & 87.9/88.0/\fbox{\textbf{87.9}}& - \\
& \multirow{1}{*}{7B} 
& 60 & 86.8/86.7/86.7 & 84.8/85.4/85.1 & 87.7/89.1/88.4 & 86.9/87.6/\textbf{87.2}  & - \\
\midrule

\multirow{1}{*}{StarCoder} & \multirow{1}{*}{16B} 
& 60 & 85.0/86.7/85.8 & 83.4/84.6/84.0 & 87.4/89.2/88.3 & 85.9/87.5/\textbf{86.7} & -  \\

\midrule

\multirow{2}{*}{CodeGen2}
& \multirow{1}{*}{16B} 
& 10 & 79.7/81.4/80.6 & 77.6/79.3/78.5 & 85.0/86.3/85.7 & 81.9/83.4/\textbf{82.7} & - \\
& \multirow{1}{*}{7B} 
& 10 & 77.7/77.3/77.5 & 76.6/77.2/76.9 & 82.2/84.0/83.1 & 79.7/80.5/80.1 & - \\

\bottomrule
\end{tabular}%
}
\end{table}

%% file: tables/rq3-callmemaybe.tex
\newcolumntype{C}{>{\centering\arraybackslash}X}
\begin{table}[t]
\caption{Comparison of different LLMs with SR on \cdata: Accuracy (\%), and Cost (\$). \mj{we need citations for all these models.}\danning{citations in table~\ref{tab:models}}
}
\renewcommand{\arraystretch}{0.9}
\label{tab:rq3-cmm}
\Large
\centering
\resizebox{0.75\columnwidth}{!}{%
\begin{tabularx}{\textwidth} {XCCCCC}

\toprule

\multicolumn{2}{c}{Approach / Model (+SR)} & K & Accuracy & Cost (\$) \\
\midrule
\callmemaybe & & - & 70.0 & - \\
\midrule

\multirow{2}{*}{GPT} & \multirow{1}{*}{4} 
& 60 &  \textbf{70.8} & 15.15\\
 & \multirow{1}{*}{3.5} 
& 40 & \textbf{73.0} & 1.01 \\

\midrule

\multirow{2}{*}{Codellama} & \multirow{1}{*}{13B} 
&  60& \fbox{\textbf{\benchmarkCMMSRSixty}}& -\\
 & \multirow{1}{*}{7B} 
& 60 & \textbf{75.3} & - \\

\midrule
deepseek-coder & 6.7B & 60 &  66.3 & - \\
\midrule

Llama3 & 8B & 60 & \textbf{71.9}  & - \\
\midrule

\multirow{2}{*}{Llama2} & \multirow{1}{*}{13B} 
& 40 & \textbf{73.0} & - \\
& \multirow{1}{*}{7B} 
& 40 & 69.7 & - \\

\midrule

\multirow{2}{*}{StarCoder2} & \multirow{1}{*}{15B} 
& 60 & \fbox{\textbf{76.4}} & - \\
& \multirow{1}{*}{7B} 
& 60 & \textbf{75.3}  & - \\

\midrule

\multirow{1}{*}{StarCoder} & \multirow{1}{*}{16B} 
& 60 & \textbf{73.0} & - \\

\midrule

\multirow{2}{*}{CodeGen2} 
& \multirow{1}{*}{16B} 
& 20 & 68.5 & -  \\
 & \multirow{1}{*}{7B} 
& 20 & 67.4 & -  \\

 \bottomrule
\end{tabularx}
}
\end{table}

%% file: future_direction.tex
\section{Challenges and Future Directions}
\label{sec:futuredirection}
\lin{some reviewers do not like mixing scientific results with ideas/discussions. Thus it is safer to keep those discussions separate from our findings sections. Plus, adding it to section 6 makes these directions only related to RQ 2. What about RQ1 and RQ3? Can we talk about more effective semantic retrieval (RQ1) as another direction?}

Our analysis of failure cases highlights several challenges, pointing to future research directions in two areas:

\paragraph{Hybrid Approaches} Root cause analysis (e.g., Section~\ref{sec:comparative_root_cause}) indicates that combining the complementary strengths of LLMs and traditional methods could improve specification generation. Hybrid approaches can harness the generalizability of LLMs along with the domain-specific precision of traditional techniques to address gaps like missing domain knowledge. Promising work has begun in this area, integrating LLMs with software testing, program analysis~\cite{lemieux2023codamosa, yuan2023no, chen2022codet}, and retrieval-augmented generation~\cite{zhang2023repocoder, ding2024crosscodeeval}.

\paragraph{Improving Prompt Effectiveness}
Improving prompts is another key direction to enhance LLM performance in specification extraction. Recent research on crafting more expressive, customizable, and domain-specific prompts~\cite{abukhalaf2023codex, beurer2023prompting, wang2022no} shows potential in guiding LLMs for better accuracy.

%% file: threats.tex
\section{Threats to Validity}
\label{sec:threats}

\paragraph{Manual Evaluation} 
To address the equivalence specification issue in \jdata (Section~\ref{sec:strudy-postprocess}), we manually evaluated results for RQ1, RQ2, and RQ4, 
and analyzed failure cases sampled in RQ3. 
To minimize biases, two authors \textit{independently} conducted evaluations with 5.9\% disagreement, resolving disagreements 
with a third author.

\paragraph{Analysis on Sampled Cases} 
In RQ3 (Section~\ref{sec:rq3}), we conduct a comparative analysis on a randomly sampled set of cases. This sampling approach could potentially limit the generalizability of our conclusions to the entire data population. To mitigate this concern, we extend the analysis to include failure cases from 2 additional models, GPT-3.5 and StarCoder (Section~\ref{sec:rq4:general}), where we observed consistent patterns, thereby reinforcing the robustness of our findings.

\paragraph{Data Leakage}
Using public datasets introduces potential risks of data leakage. To address this, we analyze performance sensitivity to the number of examples in prompts (RQ1 and RQ2) and conduct zero-shot (ZSL) and one-shot (OSL) learning experiments. ZSL shows extremely poor performance (0--0.2\% accuracy/F1), and OSL's performance is 28.9–-77.2\% lower than FSL, emphasizing the role of in-context learning.

%% file: related.tex
\section{related work}
\label{sec:related}
\todo{refer to the background section}\todo{llama}
\subsubsection{Software Specification Datasets and Extraction Methods} 
Traditional techniques for extracting software specifications from text, such as rule-based~\cite{Xie2021DocTerDF, Tan2007icommentBO, Tan2012tCommentTJ, Blasi2018TranslatingCC,Zhou2020AutomaticDA, toradocu} or ML-based methods~\cite{Lv2020RTFMAA, Tan2007icommentBO, Blasi2018TranslatingCC}, shows limited generalizability across domains and require manual effort and domain knowledge.
Our work is the first to study LLMs' capability on this task, leveraging FSL that offers improved generalizability and requires little annotated data. We evaluate LLMs on three datasets (Section~\ref{sec:study-dataset}).
Other than Jdoctor, techniques like @tcomment~\cite{Tan2012tCommentTJ}, Toradocu~\cite{toradocu}, and C2S~\cite{Zhai2020C2STN} also extract specifications from Javadoc. They are excluded from this study as C2S is unavailable and the others are outdated or less effective~\cite{Blasi2018TranslatingCC, Zhai2020C2STN}.
Advance~\cite{Lv2020RTFMAA} and DRONE~\cite{Zhou2020AutomaticDA}  are excluded
due to the absence of ground-truth specifications.

\subsubsection{Large Language Models (LLMs)}
LLMs have been developed and used for a wide range of natural language understanding tasks such as question answering~\cite{bert,roberta,T5,gpt-3,gpt-j,gpt-neo} and 
natural language generating tasks such as text summarizing~\cite{gpt-3} and machine translation~\cite{T5,gpt-3,gpt-j}. 
LLMs such as \basemodel have demonstrated their strong capabilities in numerous fields. We evaluate \nummodel state-of-the-art LLMs, varying in their design, sizes, etc., and discussed them in Section~\ref{sec:study}. While other LLMs exist, they are not explored in this study as they are unsuitable for our task~\cite{bert}, less effective~\cite{gpt-neo,gpt-j,codet5, touvron2023llama}, or unable to fit in our devices~\cite{Black2022GPTNeoX20BAO}.

\subsubsection{Applications of LLMs to SE tasks}
LLMs have also been effectively applied to SE tasks such as code completion~\cite{codegen, codellama, li2023starcoder, ding2024crosscodeeval, zhang2023repocoder}, test case generation~\cite{yuan2023no, chen2022codet, deng2023large, fsl-tester}, program repair~\cite{ ruan2024specrover, zhang2024autocoderover}, and software security~\cite{xu2023leveraging, xie2024resym,codeart}, 
often outperforming traditional methods. However, LLMs have shown limitations in areas like code summarization~\cite{sun2023automatic}, code suggestions~\cite{dakhel2023github}, and software Q\&A~\cite{xu2023we}. For example, developers frequently outperform LLMs in code-related tasks, highlighting the need for a comprehensive evaluation of LLMs in generating software specifications to identify their strengths and weaknesses in this specific SE context.

%% file: conclusion.tex
\section{Conclusion}

We present the first empirical study that 
assesses the effectiveness of \nummodel LLMs for software specifications generation. 
Our findings reveal that most LLMs achieve better or comparable performance
compared to traditional methods. Two of the 
best-performing models, \basemodel and StarCoder2-15B, as open-source models, outperform traditional approaches by 5.6 -- 10.5\% with semantically similar examples. Their strong performance makes closed-source commercial models (e.g., GPT4) less desirable due to size and cost. 
Additionally, we conduct a comprehensive failure diagnosis 
and identify the strengths and weaknesses of both traditional methods and LLMs.
The two dominant limitations of LLMs are ineffective prompts and missing domain knowledge.
Our study offers insights for future research to improve LLMs' performance on specification generation including hybrid approaches of combining traditional methods and LLMs, and improving prompts effectiveness. 

%% file: acknowledgment.tex
\section{Acknowledgement}

The authors thank the anonymous reviewers for their invaluable feedback and Guannan Wei for his input on the early draft. The research is partially supported by NSF 1901242 and 2006688 awards. It is also partially supported by the National Research Foundation grants funded by the Korean government (MSIT) (No. 2022R1A2C10911913 and 2022H1D3A2A0109297911), and the Institute of Information \& Communications Technology Planning \& Evaluation (IITP) grants funded by MSIT (No. RS-2023-00222830, RS-2024-
00337414, and 2020-0-1336).